\documentclass[aps,amsmath,amssymb,twocolumn,10pt,superscriptaddress]{revtex4-1}
\usepackage{graphicx}
\usepackage{xcolor}
\usepackage{dcolumn}
\usepackage{bm}
\usepackage[T1]{fontenc}
\usepackage[colorlinks=true,linkcolor=blue,citecolor=red,urlcolor=magenta]{hyperref}

\begin{document}

\title{Hierarchy of pairing in imbalanced three-component one-dimensional Fermi gas}
\author{Bu\u{g}ra T\"uzemen}
\email{btuzemen@ifpan.edu.pl}
\affiliation{
Institute of Physics, Polish Academy of Sciences, Al. Lotnik\'{o}w 32/46, 02-668 Warsaw, Poland
}

\author{Tomasz Sowi\'nski}
\affiliation{
Institute of Physics, Polish Academy of Sciences, Al. Lotnik\'{o}w 32/46, 02-668 Warsaw, Poland
} 

\begin{abstract}
We study a one-dimensional, three-component Fermi gas with population imbalance using the Bogoliubov-de Gennes mean-field approach. We specifically consider pairing in two channels while deliberately excluding the third by setting its interaction strength to zero. By systematically varying the interaction strength and population imbalance, we identify a rich set of spatially modulated superfluid states, including structures akin to the Larkin-Ovchinnikov phase and phase-separated domains. A clear hierarchy emerges in the pairing behavior, shaped by the competition between the local density of states and the Fermi momentum mismatch. A fidelity-based analysis of the density further distinguishes smooth crossovers from sharp spatial reorganizations. Our results shed light on how pairing symmetry and population imbalance determine the structure of spatially inhomogeneous superfluid phases in multicomponent systems with reduced dimensionality.
\end{abstract}

\maketitle

\section{Introduction}

Recent advances in the study of ultracold Fermi gases have provided a highly tunable platform for exploring exotic phases of matter~\cite{Giorgini_2008, Bloch_2008, Zwerger2011-bj}. By manipulating external fields, trapping geometries, and interaction strengths, it is possible to realize a wide variety of fermionic systems under well-controlled conditions. When a system contains multiple fermion species with distinct populations and interaction strengths, it gains additional internal structure that can support a wide variety of competing superfluid phases. This setting has been shown to host rich phase diagrams featuring unconventional pairing mechanisms, quantum phase transitions, and symmetry-breaking phenomena, including the emergence of Goldstone modes and the degeneracy of pairing manifolds in SU(N) systems~\cite{Honerkamp_2004a, Honerkamp_2004b, Rapp_2007, Capponi_2016, Cherng_2007}. Mean-field studies based on Bogoliubov-de Gennes (BdG) approaches have examined the structure of pairing fields, phase transitions, and the emergence of gapless excitations~\cite{Paananen_2006, Zhai_2007b, He_2006, Chung_2010}. 

In the two-component Fermi gas, population imbalance can result in nontrivial superfluid phases, most notably the Fulde–Ferrell–Larkin–Ovchinnikov (FFLO) state~\cite{Fulde_1964, Larkin_1965, Buzdin_2005, Bulgac_2008, Torma_2018}, where the order parameter exhibits spatial modulation, and the Sarma (interior-gap) phase~\cite{Sarma_1963, Liu_2003, Forbes_2005, Son_2006}, often accompanied by partial phase separation. In three-dimensional systems, FFLO occupies only a narrow region of the phase diagram~\cite{Castorina_2005, Mizushima_2005, Sedrakian_2005, Kinnunen_2006, Chevy_2010, Radzihovsky_2010, Gubbels_2013}. Experiments in harmonic traps with spin-imbalanced $^{6}$Li gases reveal a first-order phase separation between a balanced superfluid core and a polarized normal shell~\cite{Zwierlein_2006,Navon_2010}. Interestingly, an early contrary claim~\cite{Partridge_2006} was later traced to non-equilibrium initial conditions~\cite{Shin_2006}. While phase separation remains energetically preferred in harmonic confinement, recent studies in uniform (box-like) systems point toward LO-type stripe order—or even disordered inhomogeneous pairing—as the favoured state under comparable interaction strengths and imbalances~\cite{Radzihovsky_2010,Tuzemen_2023,Pini_2023}. 

By contrast, one-dimensional (1D) systems are more favorable for FFLO-like order. In these systems, strong quantum fluctuations suppress uniform pairing and enhance inhomogeneous superfluidity. This behavior has been confirmed using several theoretical approaches: bosonization in quasi-1D lattices~\cite{Yang_2001}, exact Bethe–ansatz combined with local density approximation in elongated traps~\cite{Orso_2007}, and density matrix renormalization group simulations of spin-imbalanced Fermi gases~\cite{Heidrich_Meisner_2010}. Experimentally, spin-imbalanced Fermi gases confined to 1D tubes also show distinctive phase structures sharply different from their 3D counterparts, as evidenced by measurements of density profiles in strong 1D confinement~\cite{Liao_2010}.

\begin{figure}
\includegraphics[width=\linewidth]{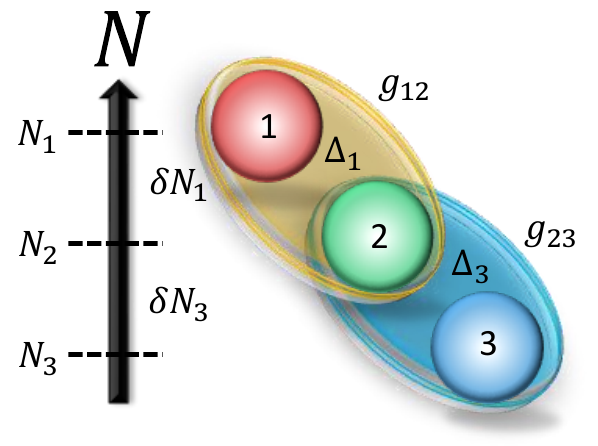}
\centering
\caption{
Schematic illustration of the system studied in this work. Three fermionic species (flavors) with unequal populations are represented by colored spheres arranged along a diagonal, with the particle number imbalance indicated by the upward arrow labeled \( N \). Pairing interactions are present between flavors 1--2 and 2--3, shown as connecting lines. No interaction is allowed between flavors 1 and 3, i.e., \( g_{13} = 0 \), while the remaining couplings are equal (\( g_{12} = g_{23} \)).
}

\label{fig:pairs}
\end{figure}

Extending the notion of imbalance to three-component systems (or other multi-flavor settings) greatly amplifies the range of possible phases. For instance, imbalanced three-component Fermi gases find natural parallels in dense quark matter, where the interplay between different quark flavors and colors creates analogous pairing competition, exhibiting pairing reminiscent of colour superconductivity in high-density QCD~\cite{Alford_1999, Alford_2001, Bowers_2002, Rajagopal_2006, Mannarelli_2006, Anglani_2014}. These systems also draw analogies with multi-band superconductors in condensed matter~\cite{Casalbuoni_2004} and connect closely to phenomena in nuclear matter, where protons, neutrons, and additional baryonic constituents form a dense, strongly interacting medium~\cite{Dean_2003}. In contrast to neutral cold atoms, QCD systems exhibit fundamentally different behavior: gluon exchange interactions enhance the parameter window supporting LO-type crystalline condensates~\cite{Leibovich_2001}, while mass differences alone can trigger the same crystalline response~\cite{Kundu_2002}. Moreover, macroscopic phase separation between color-imbalanced phases remains energetically forbidden due to the prohibitive cost of exciting long-range chromoelectric fields from bulk color charge~\cite{Steiner_2002}. Experimentally, ultracold atomic gases with multiple hyperfine states (e.g., in $^6$Li) provide an ideal experimental platform for realizing multi-component Fermi systems under precise control~\cite{Ottenstein2008, navon_2023}. 

One-dimensional systems are highly sensitive to quantum fluctuations. These fluctuations prevent the formation of true long-range order~\cite{Giamarchi_2004}. However, at zero temperature and in finite or mesoscopic systems, mean-field methods such as the BdG framework still serve an important role. They therefore remain invaluable for revealing hallmark signatures of 1D superconductors and superfluids (such as Andreev bound states and stripe- or node-forming spatial modulations) that have been captured in classic BdG studies of mesoscopic wires and ultracold gases \cite{Beenakker_1994,Mizushima_2005,Liu_2007}. Although a complete treatment of quantum fluctuations requires more advanced techniques, such as bosonization or quantum Monte Carlo, the BdG method remains a valuable tool as it reveals the dominant pairing tendencies and highlights possible inhomogeneous superfluid states.

In this work, we analyze a mesoscopic one-dimensional Fermi system composed of three distinct species with unequal populations but equal masses. Two pairing channels are active with symmetric couplings, while the third is deliberately turned off (see Fig.~\ref{fig:pairs}). Using the BdG mean-field formalism at zero temperature, we systematically examine how varying the interaction strength and population imbalance gives rise to complex superfluid configurations, including modulated and phase-separated states. We uncover a hierarchy of pairing channels set by the competition between the local density of states and the Fermi-momentum mismatch. Our analysis pinpoints the energetically preferred channel, identifies the imbalance and coupling regimes that trigger spatial modulation, and shows how these scales sculpt the resulting superfluid texture. Although quantum fluctuations inevitably play a significant role in one-dimensional systems, the BdG approach employed here effectively captures these essential pairing mechanisms at a qualitative level.

\section{The Model}

We consider a one-dimensional, three-component Fermi gas with contact interactions between fermions of different species, such as atoms in distinct hyperfine states. The system is described by the following many-body Hamiltonian:
\begin{align} \label{HamiltonianMaster}
\hat{H} = & \int\!\mathrm{d}x\sum_{i=1}^{3}  \hat{\psi}_i^\dagger(x)\left[-\frac{\hbar^2}{2m}\frac{\mathrm{d}^2}{\mathrm{d}x^2}+\sum_{j > i}g_{ij}\hat{n}_j(x) \right]\hat{\psi}_i(x) 
\end{align}
where $\hat{\psi}_i(x)$ is a fermionic field operator annihilating at position $x$ particle of component $i$, $\hat{n}_i(x)=\hat{\psi}^\dagger_i(x)\hat{\psi}_i(x)$ is the local density operator, and $g_{ij}$ is the effective strength of zero-range interactions between components $i$ and $j$.

At the mean-field level, we introduce the pairing fields:
\begin{equation}
\Delta_{ij}(x) = -g_{ij} \langle \hat{\psi}_j(x) \hat{\psi}_i(x) \rangle, \quad \text{for } i < j, 
\end{equation}
and neglect Hartree and Fock terms, retaining only the anomalous (pairing) channel. To diagonalize the resulting quadratic Hamiltonian, we perform the Bogoliubov--de Gennes (BdG) transformation of field operators:
\begin{equation}
\hat{\psi}_i(x) = \sum_n \left[ u_{n,i}(x) \hat{\gamma}_n - v_{n,i}^*(x) \hat{\gamma}_n^\dagger \right],
\end{equation}
where $u_{n,i}(x)$ and $v_{n,i}(x)$ are elements of the six-component position-dependent wavefunction
$$
\Psi_n=\{u_{n,1},  u_{n,2}, u_{n,3}, v_{n,1}, v_{n,2}, 
v_{n,3}\} ^\mathrm{T}
$$
representing quasiparticles and quasiholes in particular components. Operators $\hat{\gamma}_n$ are corresponding fermionic quasiparticle annihilation operators. Within these approximations the BdG eigenproblem for wavefunctions $\Psi_n(x)$ reads:
\begin{equation}
\mathcal{H}_\mathrm{BdG}\, {\Psi}_n(x) = \varepsilon_n \, {\Psi}_n(x),
\end{equation}
where the BdG Hamiltonian has the form:
\begin{equation}
\label{eq:ham_6by6}
    \mathcal{H}_\mathrm{BdG} =
    \left(\begin{array}{cccccc}
        h_1 & 0 & 0 & 0 & \Delta_{12} & \Delta_{13} \\
        0 & h_2 & 0 & -\Delta_{12} & 0 & \Delta_{23} \\
        0 & 0 & h_3 & -\Delta_{13} & -\Delta_{23} & 0 \\
        0 & -\Delta_{12}^* & -\Delta_{13}^* & -\,h_1 & 0 & 0 \\
        \Delta_{12}^* & 0 & -\Delta_{23}^* & 0 & -\,h_2 & 0 \\
        \Delta_{13}^* & \Delta_{23}^* & 0 & 0 & 0 & -\,h_3 
    \end{array}\right).
\end{equation}
Here $h_i = -(\hbar^2/2m) d_x^2 - \mu_i$ is the single-particle operator, and $\mu_i$ is the chemical potential used to fix the particle number.
Using components of the wavefunctions $\Psi_n$ one can define the normal $\rho_i(x)$, the kinetic $\tau_i(x)$, and the anomalous densities $\nu_{ij}(x)$ straightforwardly:
\begin{subequations}\label{eq:density_defs}
\begin{align}
\rho_{i}(x) &= \sum_{n} \bigl|v_{n,i}(x)\bigr|^2, \\
\tau_{i}(x) &= \sum_{n} \bigl|\nabla v_{n,i}(x)\bigr|^2, \\
\nu_{ij}(x) &= \sum_{n} u_{n,i}(x)\,v_{n,j}^*(x)
\end{align}
\end{subequations}
and then the pairing fields are self-consistently determined by
\begin{equation}
\Delta_{ij}(x) = -g_{ij} \nu_{ij}(x).
\end{equation}
In this notation, the densities $\rho_i(x)$ are normalized to the number of particles in respective components, $N_i = \int\!\mathrm{d}x\,\rho_i(x)$. 
The stationary configuration is obtained by minimizing the energy functional of the form
\begin{equation}
{\cal E} = \int\!\mathrm{d}x\sum_{i=1}^{3}\left[ \frac{\hbar^2}{2m}  \tau_i(x) + \sum_{j>i} g_{ij} |\nu_{ij}(x)|^2 \right].
\end{equation}

In our work, we focus only on two pairing channels by setting $g_{12} = g_{23} = g$ and $g_{13}=0$. Under these conditions, the BdG Hamiltonian \eqref{eq:ham_6by6} is simplified since it can be decoupled into two blocks acting independently on three-component wavefunctions
\begin{subequations}
\begin{align}
\varphi_n^{(1)}&=\{u_{n,1}, u_{n,3}, v_{n,2}\}^\mathrm{T}, \\
\varphi_n^{(2)} &=\{v_{n,1}, v_{n,3}, u_{n,2}\}^\mathrm{T}.
\end{align}
\end{subequations}
It turns out that due to the particle-hole symmetry of the BdG spectrum, only one of these blocks needs to be solved directly. Namely, if a function $\varphi_n^{(1)}$ is the eigenstate of the block
\begin{equation}
{\mathcal H}_\mathrm{BdG}^{(1)} =
\left(\begin{array}{ccc}
        h_1 & \phantom{-}0 & \phantom{-}\Delta_{12} \\
        0 & \phantom{-}h_3 & -\,\Delta_{23} \\
        \Delta_{12}^* & -\,\Delta_{23}^* & -\,h_2
    \end{array}\right)
\end{equation}
with a positive eigenvalue, then the negative-energy solution ($-\varepsilon_{n}$) for the other block
\begin{equation} 
{\mathcal H}_\mathrm{BdG}^{(2)} =
\left(\begin{array}{ccc}
        \phantom{-}h_1 & \phantom{-}0 & \phantom{-}\Delta_{12}^* \\
        \phantom{-}0 & \phantom{-}h_3 & -\Delta_{23}^* \\
        \phantom{-}\Delta_{12} & -\Delta_{23} & -h_2
    \end{array}\right)
\end{equation}
is directly obtained via particle-hole mapping, 
$\{\,u_1,\,u_3,\,v_2\}\;\longrightarrow\;\{\,v_1,\,v_3,\,u_2\}$.

Our particular choice of interactions recognizes the second component as the only one interacting with the other two. Thus, for further convenience, we introduce simplified notation 
for two imbalances
\begin{subequations}
\begin{align}
\delta N_1 &:= N_1 - N_2, \\
\delta N_3 &:= N_3 - N_2
\end{align}
\end{subequations}
and pairing fields
\begin{equation}
\Delta_1:= \Delta_{12}, \qquad \Delta_3 := \Delta_{23}.
\end{equation}
This approach gives us a natural measure of particle imbalances and pairing fields with respect to the mediating component.

\subsection*{Numerical Implementation}
In our approach, we solve the BdG equations iteratively on a spatial grid of $M = 200$ points, assuming a finite size of the system $L$ and imposing periodic boundary conditions. Spatial derivatives are computed using the discrete variable representation scheme~\cite{Bulgac_2013}. A convergence criterion is enforced by requiring that the changes in both the normal and anomalous densities between successive iterations fall below $10^{-9}$. We initialize the iterative procedure from a uniform configuration, then we introduce weak random noise to avoid imposing translational symmetry \textit{a priori}. Notably, if the true ground state remains uniform, the iterations naturally revert to a uniform solution despite the initial perturbation. Conversely, when the lowest-energy configuration is nonuniform for given parameters, this small symmetry-breaking seed helps the system converge to modulated density profiles. In such parameter regimes, multiple (nearly) degenerate self-consistent solutions may emerge~\cite{Tuzemen_2023}. For these scenarios, we use previously converged solutions as the initial guess to efficiently explore the various low-energy configurations.

\section{Pairing regimes under population imbalance}
Let us now investigate how pairing behavior develops in flavor-imbalanced multi-component Fermi systems. We monitor this phenomenon by analyzing the spatial profiles of the pairing fields as the interaction strength $g$ varies. While true thermodynamic phase transitions are absent in strictly 1D systems, the mean-field analysis reveals distinct regimes with qualitatively different pairing characteristics. 

\begin{figure}
\includegraphics[width=\linewidth]{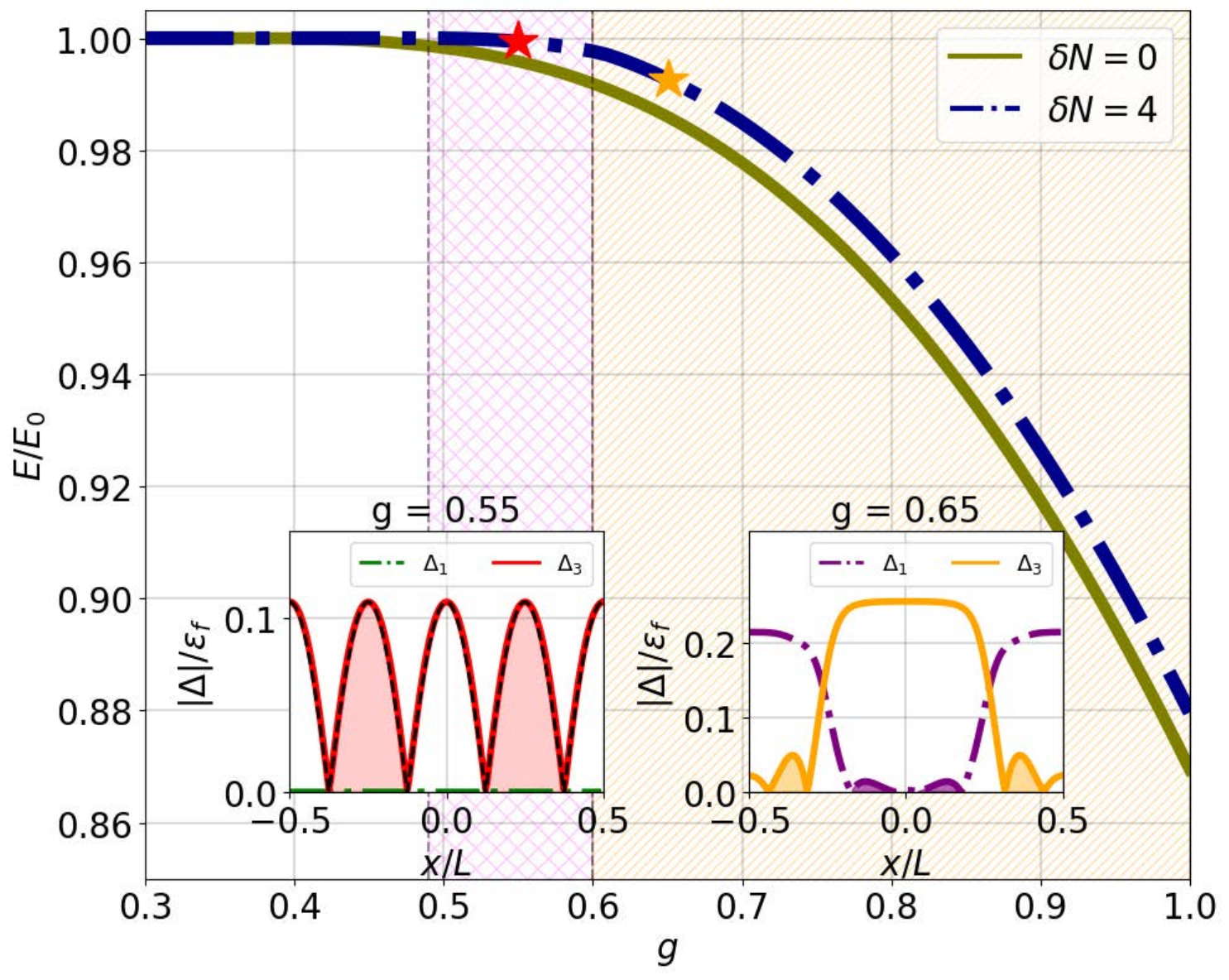}
\caption{
The total energy of the system with symmetric imbalance, normalized by the energy of a non-interacting gas as a function of the coupling constant. The solid line represents the energy of a uniform and balanced system, while the dotted line corresponds to the energy of an imbalanced system with $\delta N = 4$. The cross-hatched region ($0.49 \lesssim g \lesssim 0.60$) indicates the LO-like phase, and the diagonally hatched area ($g \gtrsim 0.60$) denotes the phase separation region. The insets show the magnitude of the pairing fields, normalized by the Fermi energy, as a function of position. The left inset corresponds to the red star in the main plot, and the right inset corresponds to the yellow star. The filled areas beneath the pairing profiles highlight regions where the phase of the pairing field is shifted by $\pi$ relative to adjacent regions.
}
\label{fig:energies}
\end{figure}

First, we inspect the properties of symmetric imbalance  $\delta N = \delta N_{1} = -\delta N_{3} $ with $N_{2} = 43$. In Fig.~\ref{fig:energies}, we show the total energy $E$, normalized by the energy of a non-interacting Fermi gas $E_{0}$, as a function of $g$ for two cases: (i) a balanced configuration ($\delta N = 0$) and (ii) an imbalanced configuration ($\delta N = 4$). We find that in the balanced case, no superfluidity appears below a certain critical interaction strength. Above this threshold, two coexisting superfluid components emerge with uniform and equal pairing fields, $\Delta_{1}(x) = \Delta_{3}(x) > 0$. By contrast, in the imbalanced system ($\delta N = 4$), when the interaction is large enough ($g\gtrsim0.49$), superfluidity initially develops in the minority pairing channel ($\Delta_{3}(x)\neq 0$), which exhibits an LO-like modulated profile. Meanwhile, the majority channel remains unpaired ($\Delta_{1}(x) \equiv 0$). Interestingly, the modulation in $\Delta_{3}(x)$ follows $|\Delta_{3}|\cos(qx)$, where $q = |k_{F,2} - k_{F,3}|$ and $k_{F,i} = \pi N_i / L$ is the Fermi momentum of a component $i$. This form is consistent with the expected Fermi surface mismatch (see the left-hand-side inset of Fig.~\ref{fig:energies}). As the interaction strength increases further ($g\gtrsim0.60$), the system enters a phase-separated regime comprising two superfluid domains, each dominated by a different pairing channel.

This hierarchy in pairing behavior originates from the fact that in one dimension, the least populated species (third component) exhibits the largest density of states (DOS) at the Fermi level. Consequently, pairing in the $\Delta_{3}$ channel is energetically favored to emerge first despite symmetric interactions and Fermi surface mismatches. In standard BCS theory, the superconducting gap $\Delta$ depends exponentially on the DOS at the Fermi level, $D(0)$, as $\Delta \sim \exp[-1/(g\,D(0))]$~\cite{Tinkham_2004}. While true long-range order is absent in 1D systems due to enhanced quantum fluctuations~\cite{Giamarchi_2004}, the local structure of the pairing field and its dependence on the low-energy DOS can still be meaningfully addressed within a mean-field framework. In particular, the BdG approach captures how spatial modulations and phase separation emerge as the low-energy DOS reorganizes with increasing interaction strength.

As the pairing field develops, the DOS-driven mechanism leads to the early dominance of $\Delta_{3}$, while the second, competing pairing channel $\Delta_{1}$ emerges only at stronger coupling. In the right-hand-side inset of Fig.~\ref{fig:energies}, the system exhibits a phase-separated configuration, with spatially distinct domains dominated by either $\Delta_{3}$ or $\Delta_{1}$. Notably, the bulk magnitude of $|\Delta_{3}|$ exceeds that of $|\Delta_{1}|$, indicating stronger pairing in the minority channel. This asymmetry once again reflects the DOS-driven pairing hierarchy. The interface between the domains is evidently not sharp. Both pairing fields extend into each other's region with damped oscillations due to the proximity effect, akin to superconductor--ferromagnet interfaces~\cite{Buzdin_2005}. These residual modulations suggest that FFLO-like features persist locally, even in the phase-separated regime. The result is a rich interplay between modulated superfluidity and domain structure.

At this point, it is worth noting that the phase diagram of the imbalanced two-component Fermi gas has been studied extensively in both 1D and 3D~\cite{Orso_2007,Guan_2007,Son_2006,Bulgac_2008,Pao_2006}. Moreover, in 1D systems, in principle, the transition from the normal phase to the FFLO state can be computed analytically~\cite{Orso_2007,Guan_2007}. It may suggest that some comparison with the corresponding transition observed in such systems can be performed. However, in our work, we focus on the mesoscopic size systems being quite far from the thermodynamic regime, having minimal imbalances (thus captured by BdG), making this comparison not straightforward and beyond the scope of this work.

\begin{figure}
\includegraphics[width=\linewidth]{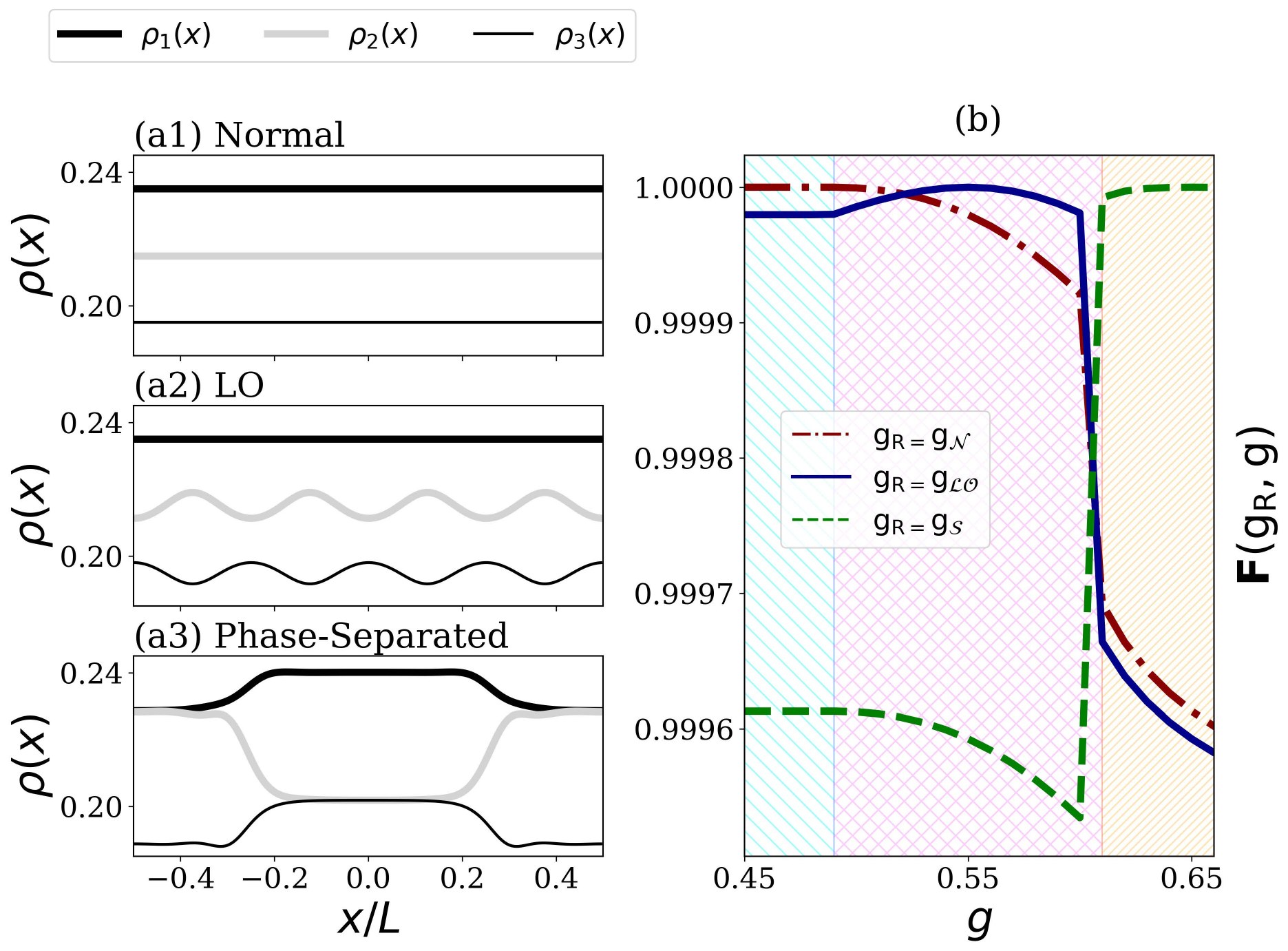}
\caption{(a1–a3) Spatial density profiles of the three components for selected coupling strengths $g_\mathcal{R} = g_\mathcal{N}$, $g_\mathcal{LO}$, and $g_\mathcal{S}$, respectively. Among the three components, only $\rho_2(x)$ (light grey) participates in both pairing channels and thus serves as a key indicator of the phase transitions. (b) The Hellinger fidelity $\mathbf{F}(g_\mathcal{R},g)$ between the reference density $\rho_2(x)$ at each $g_\mathcal{R}$ and densities at varying $g$. The three selected $g_\mathcal{R}$ values correspond to representative configurations of the $\mathcal{N}$ormal, $\mathcal{LO}$, and phase-$\mathcal{S}$eparated phases. The hatched regions in (b) mark the approximate phase boundaries.
}
\label{fig:densities}
\end{figure}

To get a better characterization of structural changes displayed in Fig.~\ref{fig:energies}, we analyze the spatial density distributions and their evolution with increasing interaction strength. In Fig.~\ref{fig:densities}a we show the particle densities $\rho_i(x)$ for three different couplings $g_\mathcal{N} = 0.45$, $g_\mathcal{LO}=0.55$, and $g_\mathcal{S}=0.65$, which represent three qualitatively distinct regimes: $\mathcal{N}$ormal, $\mathcal{LO}$-like, and with $\mathcal{S}$eparated domains. Since the second component interacts directly with both of the other species, its density profile $\rho_2(x)$ serves as a sensitive indicator of these regime changes and is marked with a colored line.

To quantify how $\rho_2(x)$ evolves with interaction strength, we first define a normalized profile $\varrho(x) = \rho_{2}(x)/N_2$ and then we employ the Hellinger fidelity as a measure of similarity between density profiles. For arbitrary two normalized profiles $\varrho_{A}(x)$ and $\varrho_{B}(x)$ this fidelity is defined as~\cite{Hellinger_1909, Matusita_1955}:
\begin{align} \label{HellingerFid}
F_H\left(\varrho_{A}, \varrho_B\right) &=
  1- \frac{1}{2}  \int\!\mathrm{d}x\,\left[\sqrt{\varrho_{A}(x)}- \sqrt{\varrho_B(x)}\right]^2 \nonumber \\
  &=\int\!\mathrm{d}x\,
    \sqrt{\varrho_{A}(x)\,\varrho_B(x)}.
\end{align}
Since in our work, the density $\varrho(x)$ depends on the interaction strength $g$ we adapt this definition by introducing the fidelity $\mathbf{F}(g_\mathcal{R},g)$ calculated as the Hellinger fidelity \eqref{HellingerFid}  calculated between the reference density obtained for a fixed reference interaction strength $g_\mathcal{R}$ (this we take as $\varrho_B(x)$) and the density obtained for varying interaction $g$ (taken as $\varrho_A(x)$). In this way, we quantify how much the density profile for an interaction g differs from a reference profile with a known pairing phase. It is clear that the fidelity $\mathbf{F}(g_\mathcal{R},g)$ defined this way is equal to $1$ only for $g=g_\mathcal{R}$ indicating full agreement of phases.
\begin{figure*}
\includegraphics[width=\linewidth]{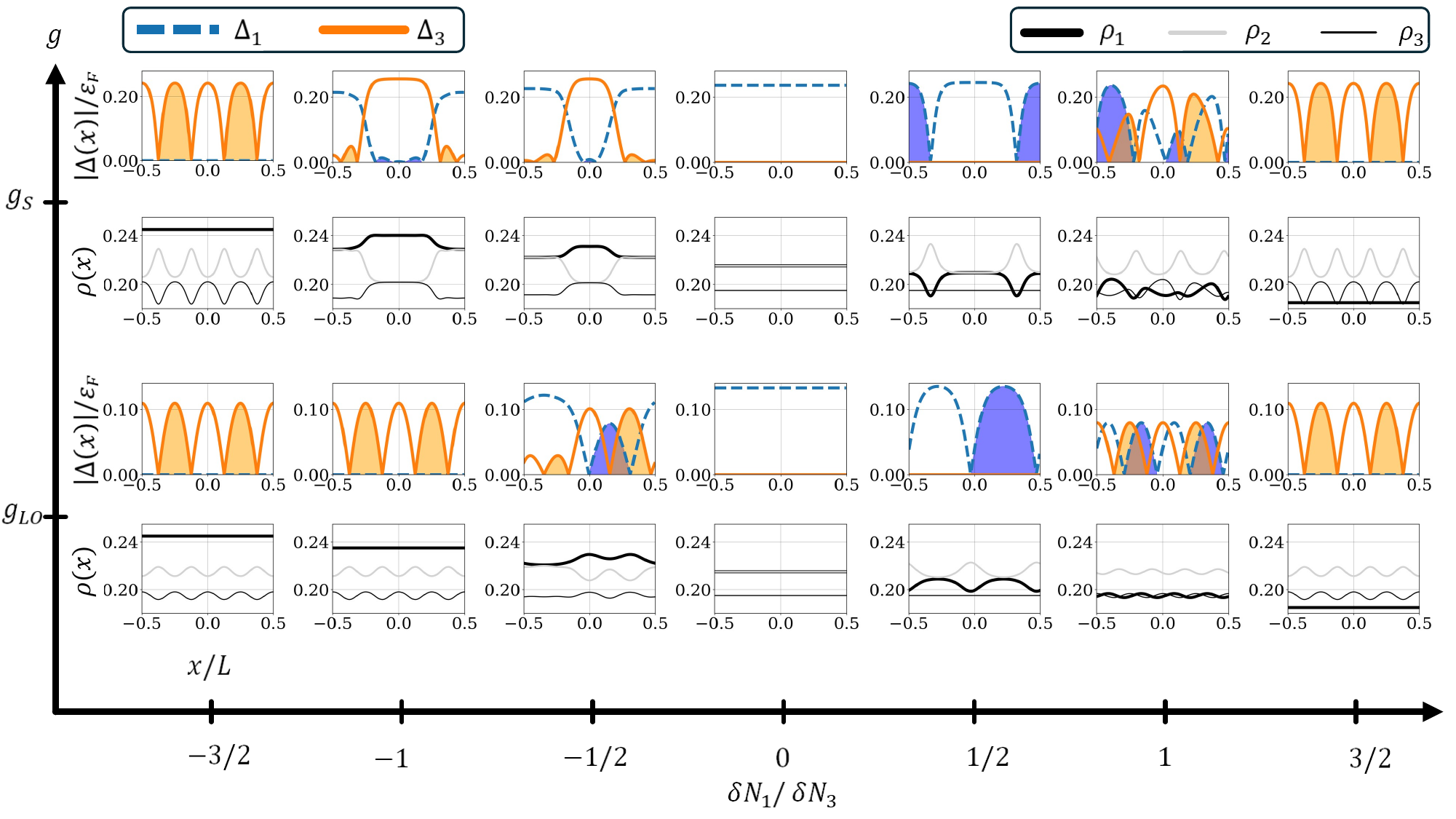}
\caption{Spatial profiles of the two non-zero pairing gaps and the three component densities in a 1-D system shown as a function of the ratio of imbalances. Dashed blue and solid orange curves show the magnitudes of the gaps $|\Delta_{1}(x)|$ and $|\Delta_{3}(x)|$, normalised by the Fermi energy $\varepsilon_{F}$. The pairing fields are accompanied by the densities $\rho_{1}(x)$ (thick black), $\rho_{2}(x)$ (light gray), and $\rho_{3}(x)$ (thin black), all as functions of the reduced position $x/L$. The upper block employs the previously shown phase-separated coupling, $g_{\mathcal{S}} = 0.65$, whereas the lower block uses the LO-branch coupling, $g_{\mathcal{LO}} = 0.55$.}\label{Fig4}
\end{figure*}

In Fig.~\ref{fig:densities}b we display three different fidelities $\mathbf{F}(g_\mathcal{R},g)$ obtained for three different reference interactions defined above, $g_\mathcal{R} \in \left\{g_\mathcal{N}, g_\mathcal{LO}, g_\mathcal{S}\right\}$. In this way, we can capture substantial transitions in the density profiles. When the normal state is taken as the reference ($g_\mathcal{R}=g_\mathcal{N}$), the fidelity decays smoothly with increasing $g$ in the LO phase, which is consistent with a gradual onset of modulation in the density $\rho_2(x)$. Then, when crossing the transition region to the separated phase, it drops rapidly. Similar behavior is observed when the LO-state is taken as reference ($g_\mathcal{R}=g_\mathcal{LO}$). Fidelity changes are rather smooth when $g$ scans over normal and LO phase, and it exhibits a rapid drop just when the system enters the phase-separated regime. This picture is fully consistent with the behavior of the fidelity calculated with respect to the separated phase ($g_\mathcal{R}=g_\mathcal{S}$). This quantity remains low at weaker couplings (normal and LO phases), and it rises steeply and saturates close to 1 when transitioning to the separated phase. This behavior confirms that the transition into the phase-separated regime involves a non-analytical change in the pairing configuration. The resulting reorganization of the spatial density profile serves as a clear signature of this structural shift.

\section{Pairing competition across imbalance}

To push our analysis further and probe the interplay between population imbalance and pairing structure, we vary the population of the target component $N_{1}$, keeping $N_{2} = 43$ and $N_{3} = 39$ ($\delta N_{3} = -4$). For convenience, we introduce the ratio of imbalances $\mathcal{R}_{N} := \delta N_{1} / \delta N_{3}$. To give a better comparison, we focus on two benchmark couplings: $g_{\mathcal{LO}}$, representative of the LO-modulated regime, and $g_{\mathcal{S}}$, characteristic of the phase-separated state. The results of this approach are collected in Fig.~\ref{Fig4}.

At moderate coupling ($g = g_\mathcal{LO}$), the system exhibits a clear hierarchy in pairing response governed by both the density of states and the Fermi momentum mismatch. For large negative asymmetries ($\mathcal{R}_{N} =  -1$ or $-3/2$), pairing occurs only in the $\Delta_{3}$ channel, which benefits from both a higher DOS and a smaller momentum mismatch. At large positive asymmetries, $\mathcal{R}_{N} =  3/2$, the DOS advantage shifts in favor of $\Delta_{1}$, but the system still prefers pairing in the $\Delta_{3}$ channel due to its smaller momentum mismatch. This highlights that momentum mismatch is dominant in selecting the pairing channel, even when it goes against the DOS. The underlying reason is that a larger mismatch requires more nodal points in the pairing field, which increases the kinetic cost and makes the state energetically less favorable. When the imbalance is present only in a single pairing channel ($\mathcal{R}_{N} =  0$), the system forms a uniform pairing in $\Delta_{1}$ while $\Delta_{3}$ is suppressed, despite its favorable DOS. Intermediate values of $\mathcal{R}_{N}$ reveal more nuanced behavior. For example, at $\mathcal{R}_{N} =  1/2$, $\Delta_{1}$ develops an LO-like structure with fewer nodes than would be required in $\Delta_{3}$, again reflecting the cost associated with large momentum mismatch. At $\mathcal{R}_{N} =  1$, both pairing channels experience the same DOS and momentum mismatch, and the system supports two coexisting LO-like states with identical modulation profiles.

The configuration at $\mathcal{R}_{N} =  -1/2$ presents a notable exception. Here, both $\Delta_{1}$ and $\Delta_{3}$ develop LO-like modulations with two and four nodes, respectively. This regime exposes a direct competition between pairing channels -- while the higher DOS supports $\Delta_{3}$, the smaller momentum mismatch favors $\Delta_{1}$. The resulting coexistence of two modulated pairings reflects a compromise between these competing factors. The spatial profiles show that the amplitude of each field is locally suppressed in regions where it overlaps with the other, highlighting the mutual influence between pairing channels.

The structure of the pairing field is directly reflected in the density profiles. When $\Delta_{1}(x) = 0$, as in the cases $\mathcal{R}_N = -3/2\,, -1\,, 3/2$, the density $\rho_1(x)$ is uniform. This reflects the absence of the pairing involving component 1. Conversely, the densities $\rho_2(x)$ and $\rho_3(x)$ exhibit modulations that follow the structure of the active pairing field $\Delta_{3}(x)$, {\it i.e.}, the density dips and peaks align with the nodes and antinodes of the pairing amplitude. At $\mathcal{R}_N=1/2$ the roles reverse. The density of the third component $\rho_{3}(x)$ stays uniform and the pairing field $\Delta_{3}(x)=0$. This is accompanied by modulations of densities $\rho_{1}(x)$ and $\rho_{2}(x)$ which are influenced by pairing field $\Delta_{1}(x)$. When the imbalance involves only two species, $\mathcal{R}_N=0$, all three densities remain essentially flat, which is consistent with the absence of spatial modulation. Finally, the intermediate ratios $\mathcal{R}_N=-1/2$ and $1$ allow both pairing channels to survive. Then, their competition produces two LO-like waves with different wavelengths, so every density profile shows a pronounced, commensurate modulation.

As the interaction strength increases ($g = g_\mathcal{S}$), similar patterns persist, but signatures of phase separation become more prominent, particularly for $\mathcal{R}_{N} =  -1/2$ and $-1$. The node counts remain consistent with momentum mismatch, and pairing fields show mutual suppression in overlapping regions. For $\mathcal{R}_{N} =  1/2$, the profile of $\Delta_{1}$ departs from the expected cosine form. It develops a flat-top structure, possibly due to nonlinearities in the self-consistent BdG equations or indirect competition between the channels. Finally, at $\mathcal{R}_{N} =  1$ we find that the solution becomes numerically unstable and the profiles appear irregular. This, of course, does not necessarily mean there is no stable solution. It rather reveals that in the case of imbalanced systems, the energy landscape near the actual ground state of the system is highly quasi-degenerated and the numerical algorithm of minimization is not efficient. In consequence, it often gets stuck in a local minimum. This problem, pointing to some limitations of the methodology applied, has already been captured in previous studies~\cite{Tuzemen_2023}.

\section{Conclusion}

We have performed a detailed mean-field study of a one-dimensional Fermi system composed of three distinct species with imbalanced populations. By focusing on two strongly coupled pairing channels, we identified a range of nontrivial superfluid structures, including LO-like modulated states, phase-separated domains, and regimes where both pairing channels coexist in a spatially intertwined manner. These configurations arise from the competition between the density of states advantage of the minority species and the momentum mismatch between Fermi surfaces, giving rise to a rich and highly responsive superfluid landscape.

We established a clear hierarchy of pairing channels governed by the enhanced density of states in the least-populated component and the nodal constraints imposed by the Fermi momentum mismatch. We found that one pairing channel typically dominates whenever it can minimize the formation of nodes, even if another channel possesses a more favorable density of states. At moderate coupling strengths, this competition yields exotic LO-like configurations, where both channels partially modulate the superfluid, mutually suppressing each other's amplitudes in overlapping regions. At stronger couplings, the system transitions into a phase-separated regime, characterized by domain-like structures in which each pairing channel occupies a distinct spatial region. Notably, remnants of the LO modulation persist locally within these domains.

The fidelity-based analysis of the density profiles confirms the presence of multiple qualitative regimes. Small but continuous changes in the density distribution characterize the normal-to-LO crossover, whereas a more abrupt reorganization of the spatial profiles marks the onset of phase separation. We have demonstrated that mean-field theory, although neglecting quantum fluctuations, still provides valuable insights into the spatially inhomogeneous pairing patterns that arise under mesoscopic conditions.

\section*{Acknowledgements}
This research was supported by the National Science Centre (NCN, Poland) within the OPUS project No. 2023/49/B/ST2/03744. For the purpose of Open Access, the authors have applied a CC-BY public copyright licence to any Author Accepted Manuscript version arising from this submission.

\section*{Data availability}
The data that support the findings of this article are openly available~\cite{zenodo}.

\bibliography{biblio}

\end{document}